\documentclass[preprint]{revtex4-2}
\usepackage[utf8]{inputenc}
\usepackage{color,soul}
\usepackage{hyperref}
\usepackage{graphicx}
\usepackage{comment,xcolor}
\usepackage{mathtools}
\usepackage{amsmath}
\usepackage{amssymb}
\usepackage{natbib}
\usepackage{multirow}
\usepackage[colorinlistoftodos]{todonotes}

\newcommand{\desy}{{DESY, Platanenallee 6, D-15738 Zeuthen, Germany}}

\newcommand{\huberlin}{{Humboldt-Universit\"at zu Berlin, Unter den Linden 6, D-10099 Berlin, Germany}}

\newcommand{\mcgill}{{Department of Physics, McGill University, Montreal, Qu\'ebec H3A 2T8, Canada}}

\newcommand{\com}{center of mass}

\newcommand{\prap}{pseudorapidity}

\begin{document}

\title{Probing the Cosmic Ray Background of Gamma-Ray Astronomy with Hadron Colliders}

\author{Clara E. Leitgeb}
\affiliation{\huberlin}
\email{clara.leitgeb@physik.hu-berlin.de}

\author{Andrew M. Taylor}
\affiliation{\desy}
\email{andrew.taylor@desy.de}

\author{Robert D. Parsons}
\affiliation{\huberlin}
\email{daniel.parsons@physik.hu-berlin.de}

\author{Kenneth J. Ragan}
\affiliation{\mcgill}
\affiliation{\desy}
\email{ragan@physics.mcgill.ca}

\author{David Berge}
\affiliation{\desy}
\affiliation{\huberlin}
\email{david.berge@desy.de}

\author{Cigdem Issever}
\affiliation{\desy}
\affiliation{\huberlin}
\email{isseverc@physik.hu-berlin.de}

\begin{abstract}
Hadronic cosmic particles (cosmic rays) and gamma rays are constantly absorbed in the Earth's atmosphere and result in air showers of secondary particles. Cherenkov radiation from these atmospheric events is used to measure cosmic gamma rays with ground-based telescopes. We focus here on the dominant hadronic cosmic-ray-initiated background events in the atmosphere, which give rise to gamma-ray like air showers for gamma-ray telescopes. It is shown that only a small subset of hadronic cosmic-ray interactions, those which produce a large energy neutral pion, are responsible for this background. We subsequently address how the predictions of this background vary depending on the hadronic interaction model adopted. The \prap\, range of the energetic pions, with respect to the shower axis produced in these background events, is shown to be large. We show that collider experiments, specifically LHCf and RHICf, probe cosmic ray interactions precisely within this \prap\, range. Present and future measurements with these instruments are shown to be able to test the ability for current hadronic interaction models to accurately describe these background events.
\end{abstract}

\maketitle

\section{Introduction}
\label{intro}


Ground-based gamma-ray telescopes, operating in the very high energy (VHE; $E>$100~GeV) range, image cosmic gamma rays by utilizing the Earth's atmosphere as a target for the interaction of these gamma rays. The identification of the gamma-ray initiated electromagnetic air shower is a key component of this detection technique. The products of these air showers can be detected either directly at ground level by instruments such as HAWC~\cite{Albert_2020} and LHAASO~\cite{Cao_2024}, or by observation of the Cherenkov emission produced by the air shower throughout the atmospheric air column, by imaging atmospheric Cherenkov telescopes
(IACTs) such as H.E.S.S., MAGIC, and VERITAS (see \cite{2015arXiv151005675H} and references therein).
By utilizing the atmosphere as part of the detector, these instruments achieve very large effective areas (more than 10$^5$\,m$^2$ for IACTs), but require proper knowledge of both the atmosphere as well as the physics of the air shower development.


In the case of gamma-ray-initiated air showers, the particle physics of the electromagnetic interactions primarily involved in the
showers can be accurately modeled by quantum electrodynamics, for example by the EGS4 model \cite{Nelson:1985ec,Nelson_1990}. However, as can be derived from the Crab nebula and cosmic ray (CR) fluxes given in \cite{2020NatAs...4..167H} and \cite{ParticleDataGroup:2024cfk}, the gamma-ray showers 
of interest for typical sources are outnumbered by at least a factor of 10$^3$ by hadronic CR-initiated air showers. These CR particles, composed predominantly of protons for the energies of interest, continuously bombard the Earth's atmosphere where they deposit their energy in the form of hadronic air showers. These hadronic air showers, dominated by pion production and subsequent particle cascades \cite{MATTHEWS2005387}, must be efficiently rejected in order for even the brightest gamma-ray sources to be detectable \cite{Crab1987}.


In the analysis of IACT data, the selection of gamma-ray-like events is performed by measuring the differences in shower shape, as measured through the Cherenkov light they produce, taking advantage of the inherent differences in shower development between gamma-ray-induced and hadron-induced events. On average, cosmic-ray hadrons produce air showers that are longer, wider, and have more sub-structure than equivalent gamma-ray-induced showers \cite{2015arXiv151005675H}, allowing the rejection of the CR-initiated background at around the 99\% level. However, even with such efficient background rejection, due to the large initial CR rates, a significant number of misidentified hadron-induced background showers remain. Even if rarely, CR showers can mimic gamma-ray shower behavior and constitute an irreducible background, the level of which must be understood to accurately measure VHE gamma-ray fluxes from sources.


To account for this background and subtract it, one can either employ data-driven techniques~\cite{2007A&A...466.1219B} or use Monte Carlo simulations~\cite{2008PhRvL.101z1104A}. For the latter, a good understanding of both the underlying hadronic CR flux and the interaction physics describing how these hadrons interact with the atmospheric gas is required. Measurements of the CR flux as a function of energy (ie. the CR spectrum) at the TeV energy scale of interest here have been made by several space-based detectors. Specifically, AMS-02 \cite{2015PhRvL.114q1103A,PhysRevLett.113.221102}, CALET \cite{PhysRevLett.122.181102,PhysRevLett.119.181101}, and DAMPE \cite{2019SciA....5.3793A,2017Natur.552...63D} have all made recent measurements of the CR energy spectrum. The differences in the CR energy spectrum measurements from these different instruments are at the $\sim$10\% level. However, the difference in the hadronic interaction model predictions are larger than this, and thus limit the precision of the background estimate. Attention should, therefore, be focused on the uncertainty introduced by the hadronic interaction model used to describe CR interactions with the atmosphere.

With regards to interaction physics, air shower interactions are described by phenomenological models describing the cross sections and final state particle distributions. The free parameters of several of these models have now been tuned to particle interaction data from the Large Hadron Collider (LHC), which has given rise to a suite of post-LHC interaction models, including 
QGSJET-II.04 \cite{Ostapchenko_2011}, 
EPOS-LHC \cite{EPOSLHC}, 
and SIBYLL2.3d \cite{Sibyll23d}. 
CORSIKA\cite{Heck_1998} is a widely-used framework for the simulation of air shower physics; the current version of CORSIKA, CORSIKA~8~\cite{huege2022corsika}, embeds the three aforementioned interaction models, as well as others (Pythia \cite{Pythia83}, UrQMD \cite{UrQMD}, PROPOSAL \cite{2020JPhCS1690a2021A}), and the addition of FLUKA \cite{FLUKA} is planned. 
However, the interaction phase space probed by the LHC experiments, used for the tuning of these interaction models, is different to the phase-space region of relevance to IACT instruments (and focused on in this work). We note that here, and throughout the paper, we use the term phase space to refer to the energy and \prap\ (ie. the angle of the secondary particle momentum from the beam axis) distribution of secondary particles. Specifically, the experiments used so far for this model tuning have probed relatively low energy-exchange fractions between the parent protons and the pions produced, as well as relatively low \prap\ values. 
    

Further insight from particle collider experiments, testing CR interaction models in the part of the interaction phase space giving rise to hadronic air showers that appear gamma-ray like, is provided by dedicated forward-focusing experiments such as LHCf \cite{LHCf_TDR}. The LHCf experiment consists of two calorimeter arms located at a distance of approximately $140\,\text{m}$ along the LHC beam line in either direction from the ATLAS interaction point. Serendipitously, the LHCf fiducial range for $\pi^{0}$ detection covers precisely the interaction phase space range, matching both the high energy-exchange fractions from the beam particles as well as the small forward angles (ie. high pseudorapidities), that are of relevance for CR showers misidentified as gamma rays (as further discussed in Sec.\,\ref{pseudorapdity_range} below). Furthermore, the planned proton-oxygen collision run at the LHC is of strong interest for improving CR shower modeling. Consequently, the main focus of this study will be on proton-oxygen and proton-proton collisions due to the availability of the corresponding data.

In this paper, we investigate the phase-space region of relevance for the hadronic cosmic-ray atmospheric interactions which dominate the contribution to gamma-ray-like hadronic cascades for IACTs, with an aim to determining how interactions in this region can be probed by present and future particle collider experiments. In Sec.~\ref{Air_Showers}, we develop a simplified model for the rejection of background by IACTs, highlighting the phase space region for which the energy exchange fraction of the incoming beam particle (the cosmic ray) to a $\pi^{0}$ particle, and subsequently to electromagnetic particles, is the dominant energy flow. In Sec.~\ref{collider_experiments} the interaction parameter space probed by current LHC and other collider experiments is determined, highlighting which experiments best probe the interaction phase space of relevance for IACTs. In Sec.~\ref{LHCf_Comparison}, a comparison of the interaction models to LHCf data is made. In Sec.\ref{Discussion}, we discuss the importance of upcoming future runs by LHCf, which will further probe this interaction phase space. 

\section{Atmospheric Air Shower Simulation}
\label{Air_Showers}

High-energy gamma-ray-initiated air showers and a subset of high-energy CR air showers can have remarkably similar shower development characteristics. Specifically, it has been shown that hadronic showers for which the first inelastic interaction is a low multiplicity event producing a high-energy $\pi^{0}$ look like gamma rays for IACTs~\cite{2007APh....28...72M,2018APh....97....1S}. In order to quantify the fraction of energy exchanged into the electromagnetic channel through this initial interaction in the atmosphere, we carry out simulations of high-energy air showers initiated by proton-oxygen ($pO$) interactions using the air shower simulation tool CORSIKA (version 76900) \cite{CORSIKA} and a simplified model of a Cherenkov telescope array (described in detail in Appendix \ref{Appendix1}).

In these CORSIKA simulations, proton induced air showers were initiated at a fixed first interaction of the proton showers at 17.55\,km altitude \cite{pastor-gutierrez_sub-tev_2021} (and the initial interaction products observed 1\,cm below this) and the target nucleus in the atmosphere fixed to oxygen. In total 5$\times10^5$ vertical proton showers were simulated at an energy of 1~TeV, subsequently recording particles and Cherenkov photons both directly below the interaction point and lower in the atmosphere at an altitude of 1800\,m above sea level.

Although nitrogen is more abundant in the atmosphere than oxygen, we use oxygen as reference in light of the planned proton-oxygen collision run at the LHC. The nuclear number of the most prevalent isotopes of nitrogen and oxygen in the atmosphere differs by two. This is not expected to cause large differences in the predictions for CR showers, especially compared to the model discrepancies we are investigating in the following.


\begin{figure*}
    \centering
    \includegraphics[width=0.6\linewidth]{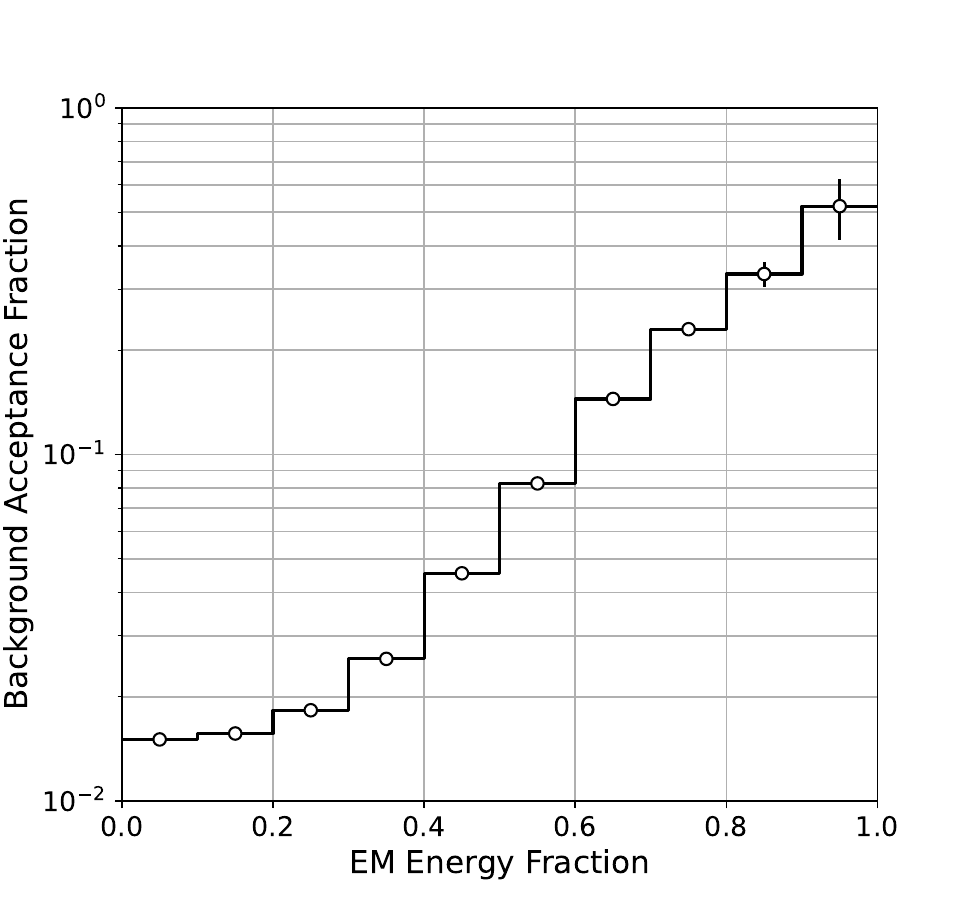}
    \caption{Fraction of background proton events simulated with Sibyll 2.3c passing gamma-ray likeness cuts using our simplified description for an array of Cherenkov telescopes (see  Appendix~\ref{Appendix1}). The background fraction is shown as a function of the fraction of the initial proton energy in the EM channel (electrons, positrons, photons and neutral pions) 1\,cm below the first interaction point.}
    \label{fig:hadronic-gammas}
\end{figure*}

Fig.~\ref{fig:hadronic-gammas} shows the fraction of 1\,TeV cosmic-ray air-shower events passing background rejection cuts (defined in Appendix \ref{Appendix1}) as a function of the electromagnetic (EM) energy fraction immediately after the first interaction point. This EM fraction, consisting of the contributions of photons, electrons, and positrons, is dominated by neutral pions and their decay products. Hence, in the following sections, we use the energy fraction of neutral pions produced in the primary interaction as a proxy for the EM energy fraction of the shower. At EM fractions below $\sim$30\%, strong background rejection is seen to be possible, with only 1-2\% of events passing the cuts. However, above 40\% EM fraction, a reduction in rejection power is seen with increasing EM fraction. By an EM fraction value of 70\% more than 15\% of events pass selection cuts. This enhancement at high EM fraction shows that significant energy in the EM channel enhances the ability of an air shower to appear gamma-ray like.

\subsection{Hadronic Interaction Simulation}
\label{Simulation}

Deeper insight into the nature of the subset of the hadronic interactions which give rise to a large electromagnetic shower component can be provided through further Monte Carlo simulation studies. We carry out these simulations, for the case of proton-proton ($pp$) interactions, using the Pythia 8.310, EPOS-LHC \cite{EPOSLHC}, QGSJET II-04 \cite{Ostapchenko_2011} and Sibyll 2.3d \cite{Sibyll23d, Sibyll21} interaction models. For the Pythia simulation, the parameters are set according to the ATLAS A3 tune \cite{ATLAS_A3_PythiaTune} as well as the values tuned to forward particle spectra in LHCf (``forward tune'') \cite{Forward_PythiaTune}. For EPOS-LHC, QGSJET II-04 and Sibyll 2.3d, the implementations provided by the CRMC package (version 1.8.0) \cite{CRMCpackage} were used. No detector simulation is applied.

\begin{figure*}
\centering
\includegraphics[width=0.8\linewidth]{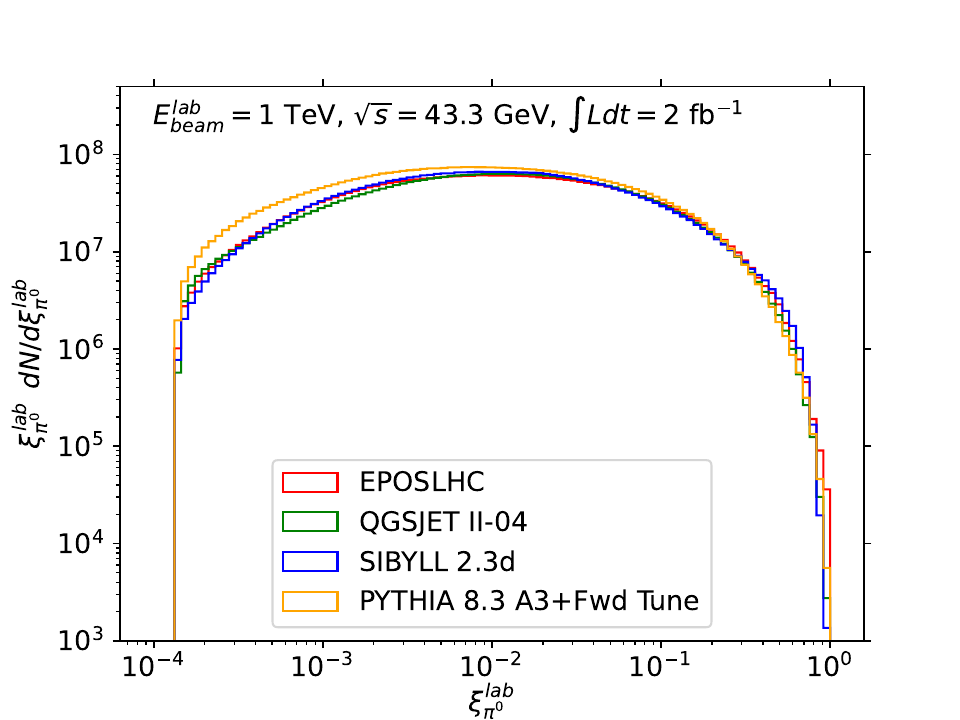}\\
\includegraphics[width=0.8\linewidth]{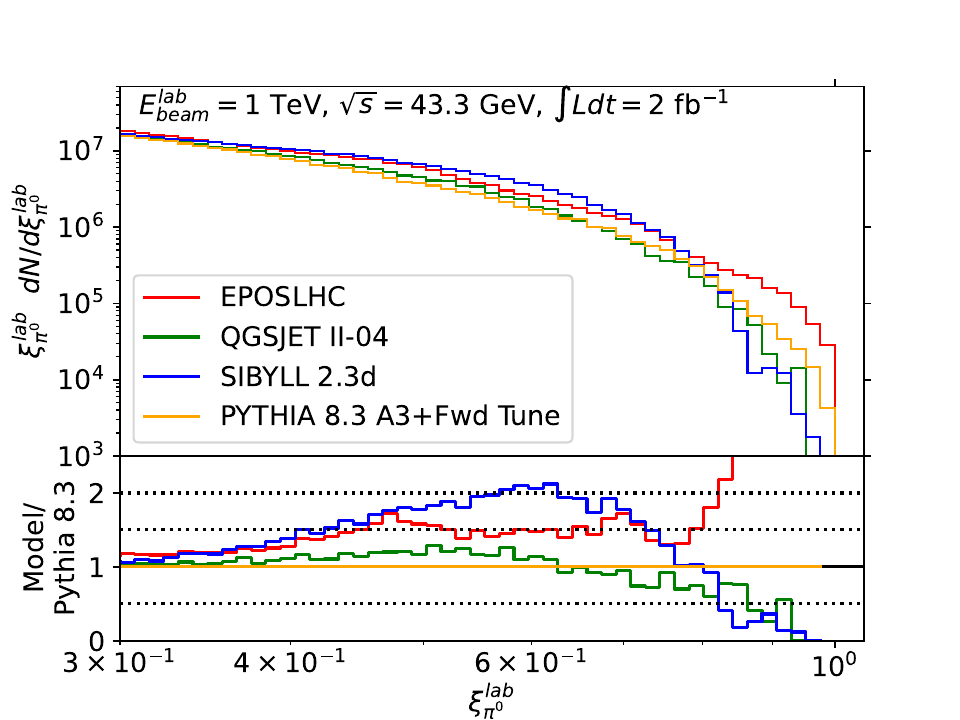}
\caption{{\bf Top}: The predicted $\pi^{0}$ energy spectrum produced by a proton beam energy of $1\,\text{TeV}$, $\xi_{\pi^{0}}^{\text{lab}}= E_{\pi^0}/E_{\rm beam}$, through collisions with target protons at rest, for four different interaction models. Each $\pi^{0}$ produced in a collision corresponds to a separate entry in the histogram.\\
{\bf Bottom}: Same as above, but zoomed in to the cutoff region in the final energy decade. All quantities in both plots are evaluated in the lab frame.} 
\label{fig:pi0spec}
\end{figure*}

Fig.~\ref{fig:pi0spec} shows the predictions for the $\pi^{0}$ energy spectra, produced through $pp$ collisions, for each of the hadronic interaction models considered, represented by the variable
\begin{equation}
\xi^{\text{lab}}_{{\pi}^{0}} = \frac{E^{\text{lab}}_{\pi^{0}}}{E^{\text{lab}}_{\rm beam}}.
\end{equation}
In the figure, logarithmic $\xi^{\text{lab}}_{{\pi}^{0}}$ bins have been adopted. We use here and in the following the ``lab'' superscript to denote the frame in which one of the two collision particles is at rest. As seen in this figure, the models are noted to roughly agree in the central region of the plot, responsible for determining the total $\pi^{0}$ production cross section, 
\begin{eqnarray}
\sigma_{\text{total}} = \frac{1}{\mathcal{L}}\int_{-\infty}^{0}\xi_{\pi^{0}}^{\rm lab}(dN/d\xi_{{\pi}^{0}}^{\rm lab})d\ln \xi_{\pi^{0}}^{\rm lab},
\end{eqnarray}
where $\mathcal{L}$ is the integrated luminosity. In Fig.~\ref{fig:pi0spec} this is set to $2\,\text{fb}^{-1}$.

The contribution to this integral for the total cross section is dominated by the central region of the integration range, around an energy transfer value of $\xi_{{\pi}^{0}}^{\rm lab}\approx 10^{-2}$. In contrast, the distribution functions for the hadronic models diverge towards both the low and high energy transfer end of the spectrum. The lower cutoff in the spectrum is kinematic in origin, resulting from the production of neutral pions at rest in the \com\ frame. Of most relevance for this work is the difference between the models in the large $\xi_{\pi^{0}}^{\rm lab}$ region, where the $\pi^{0}$ particle carries a large fraction, $\xi_{\pi^{0}}^{\rm lab}>0.1$, of the original beam's energy. The lower panel in Fig.~\ref{fig:pi0spec} shows a zoom-in on this part of the neutral pion spectrum. The hadronic interactions taking part in this range contribute directly to the irreducible cosmic-ray background for gamma-ray induced air showers.

\section{Relevant Collider Experiments for Cherenkov Telescopes}
\label{collider_experiments}

\subsection{Relevant pseudorapidity range of \texorpdfstring{$\pi^{0}$}{pi0} mesons in \texorpdfstring{$pp$}{pp} collisions}
\label{pseudorapdity_range}

\noindent
The dominant background for gamma-ray initiated air shower events observed by IACTs is constituted by hadronic interaction events where a single ``forward" $\pi^{0}$ meson carries a large fraction of the original beam energy. Here we adopt the ad-hoc definition commonly used in the community
\cite{Ohishi:2021hfu}, namely that these events are characterized by events in which the produced $\pi^{0}$ carry more than $70\%$ of the initial beam energy, namely $\xi^{\text{lab}}_{{\pi}^{0}} > 0.70$. 

To improve the predictive power that hadronic interaction models provide for this subclass of $pp$ interaction events, the corresponding phase space in existing hadron collider experiments has to be identified and probed through measurements. The reference CR scenario - namely, a proton beam of $1\,\text{TeV}$ colliding with a fixed proton target - corresponds to a \com\ energy of about $43.3\,\text{GeV}$, significantly below LHC collision energies. In order to compare the collision kinematics between these very different energies, it is convenient to use the \prap\ variable, $\eta$.

The \prap\ of a particle with energy $E$, momentum $p$ and longitudinal momentum (parallel to the beam axis) $p_{z}$, which is produced in a collision of two proton beams, each with energy $E_{\text{beam}}$, is defined as
\begin{align}
\eta &= \frac{1}{2}\, \ln\bigg(\frac{|p| + p_{z}}{|p| - p_{z}}\bigg) \approx \frac{1}{2}\,\ln\bigg(\frac{(E+p_{z})^{2}}{E^{2}-p_{z}^{2}}\bigg).
\end{align}
Considering a ``maximal'' case, in which a produced $\pi^{0}$ meson inherits the entire proton energy ({\it i.e.} $E_{\pi^{0}}=E_{\text{beam}}$ and $p_{{\text{T}},\pi^{0}}\rightarrow 0$, where $p_{{\text{T}},\pi^{0}}$ is the transverse momentum component of the $\pi^{0}$), the upper bound for the $\pi^{0}$ \prap\ range is given by
\begin{align}
\eta_{\text{max}} \approx \frac{1}{2}\,\ln\bigg(\frac{4E_{\text{beam}}^{2}}{m_{\pi}^{2}}\bigg)=\ln\bigg(\frac{\sqrt{s}}{m_{\pi}}\bigg).
\label{eqn:etamax}
\end{align} 
Taking the case in which the $\pi^{0}$ meson carries $70\%$ of the beam energy as the lower bound of the $\pi^{0}$ energy range of interest, the corresponding lower edge of the \prap\ region is given by
\begin{align}
\eta_{\text{min}} \approx \frac{1}{2}\,\ln\bigg(\frac{0.7^{2}\, s}{m_{\pi}^{2} + p_{\text{T}}^{2}}\bigg) = \ln\bigg(\frac{0.7\sqrt{s}}{\sqrt{m_{\pi}^{2} + p_{\text{T}}^{2}}}\bigg).
\label{eqn:etamin}
\end{align}
Simulations with Pythia 8.3, EPOS-LHC, QGSJET II.04 and Sibyll 2.3d show that the maximum transverse momentum of a $\pi^{0}$ meson carrying a $70\%$ fraction of the beam energy is approximately $1.5\,\text{GeV}$, with only small variation with \com\ energy.

\begin{figure*}
\centering
    \includegraphics[scale=0.5]{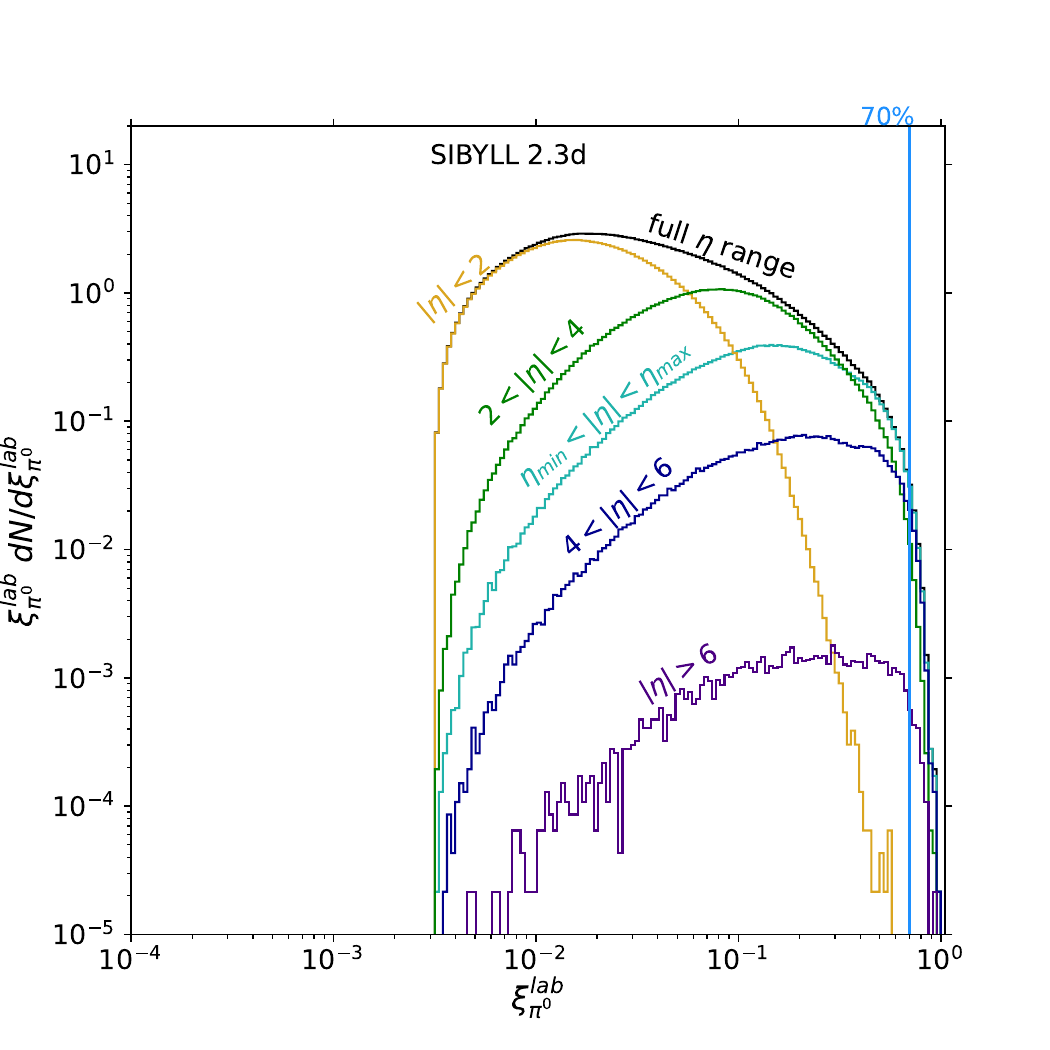}\\
    \includegraphics[scale=0.5]{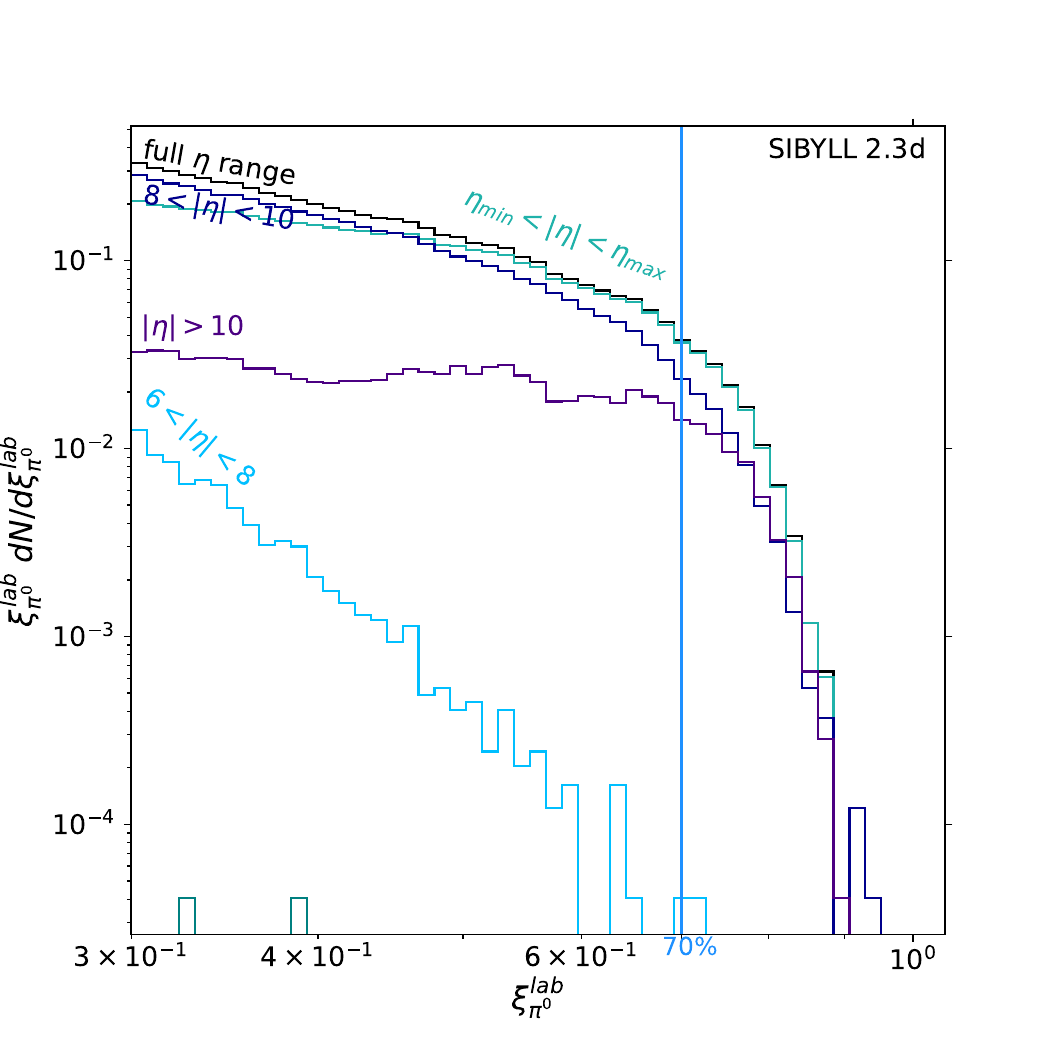}
\caption{The $\pi^{0}$ energy spectrum produced through the collision of a $1\,\text{TeV}$ proton beam with a fixed proton target (corresponding to a \com\, of $43.3\,\text{GeV}$). The full spectrum is decomposed into different \prap\ ranges (evaluated in the \com\ frame). The upper plot shows the full spectrum and the lower plot shows the result for the zoomed-in, high energy fraction region. Depicted here are the predictions according to the Sibyll 2.3d event generator. The \prap\ range defined by equations~\ref{eqn:etamax} and ~\ref{eqn:etamin} is shown in turquoise and corresponds to $3.0<|\eta_{\pi^{0}}^{\text{COM}}|<5.77$. The 70\% percent energy fraction benchmark point is denoted by the blue vertical line.}
\label{fig:breakdown}
\end{figure*}

In Fig.~\ref{fig:breakdown} a decomposition of the differential neutral pion spectrum, as a function of the ratio of the neutral pion energy to the original proton energy, into \prap\ bins is shown for all considered interaction models. The spectrum is different from the one displayed in Fig.~\ref{fig:pi0spec} in that the events are generated in the \com\ frame and the pion energy in the CR collision frame (``lab frame'') is obtained by applying a Lorentz boost in either positive or negative beam direction. Because of the symmetry in the \com\ frame in $pp$ collisions, the direction of the Lorentz boost is ambiguous, with either the positive or negative z-direction being possible. To break this ambiguity, we choose the direction for which the maximum pion energy is obtained, on an event-by-event basis. Therefore the energy range is shifted towards the high end of the spectrum in Fig.~\ref{fig:breakdown} with respect to Fig.~\ref{fig:pi0spec}. The higher edge of the energy spectrum ($>70\%$ of the beam energy) is dominated by contributions from neutral pions in the \prap\ bin bounded by Eqn.~\ref{eqn:etamax} and Eqn.~\ref{eqn:etamin} for all of our four canonical interaction models. For a \com\ energy of $43.3\,\text{GeV}$ the derived \prap\ range is given by $3.0<|\eta_{\pi^{0}}^{\text{COM}}| < 5.77$ (see the turquoise curve in the lower panel of Fig.~\ref{fig:breakdown}). The coverage of the $\xi_{\pi^{0}}^{\text{lab}}>70\%$ region of the spectrum by the \prap\ region defined by Eqn.~\ref{eqn:etamin} and Eqn.~\ref{eqn:etamax} is also validated for higher collision energies, shown in Sec.~\ref{appendix2} in the appendix.

\section{Phase-Space Coverage of Collider Experiments}
\label{LHCf_Comparison}

\begin{figure}
\centering
\includegraphics[scale=1]{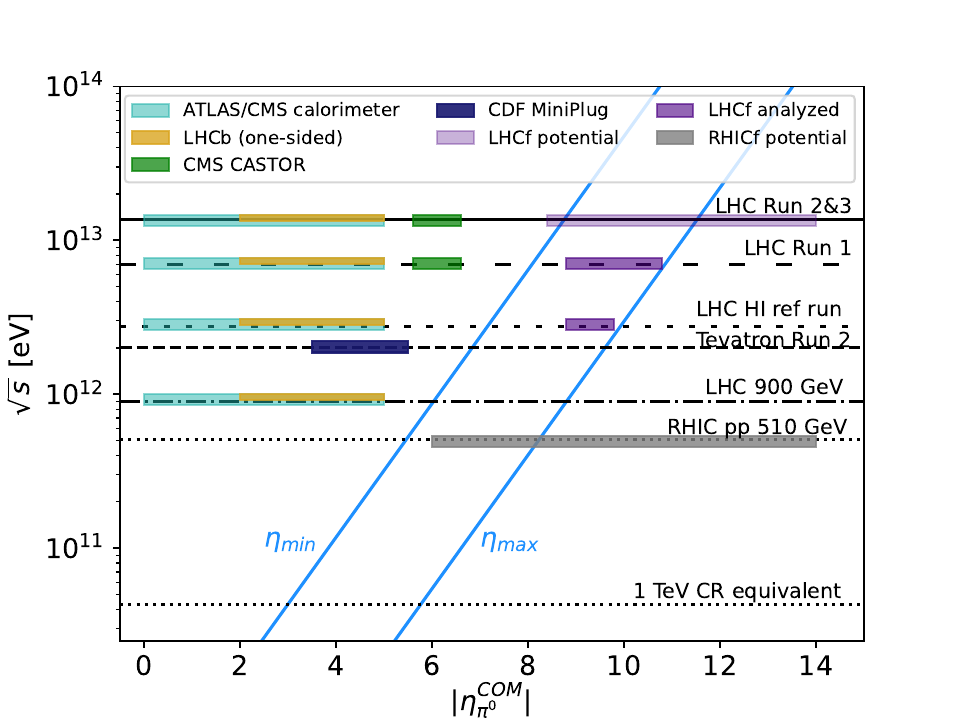}
\caption{Relevant $|\eta_{\pi^{0}}^{\text{COM}}|$ region in the \com\ frame for gamma-ray like air showers from cosmic rays (light blue lines). The black horizontal lines correspond to available data in terms of \com\ energy and the colored boxes indicate the $|\eta|$ coverage of the corresponding experiments. The label ``potential'' refers to LHCf and RHICf data that has already been recorded but not published.}
\label{fig:EtaRegions}
\end{figure}

After identifying the \prap\, range of interest, which depends upon the \com\ energy for the collision, the next step is to identify existing or future experiments that can reconstruct and identify $\pi^{0}$ mesons in that phase-space region.

At the LHC, two proton beams are collided at a variety of different \com\ energies. During Run 1, the LHC was operated at $7\,\text{TeV}$ and $8\,\text{TeV}$, while the collision energy was increased to $13\,\text{TeV}$ and $13.6\,\text{TeV}$ for Run 2 and Run 3, respectively. During commissioning periods, data was also taken under special running conditions with a \com\ energy of $900\,\text{GeV}$. Similarly, there are dedicated $pp$ runs at specific \com\ energies that are used as reference for heavy ion collision runs. An example is the heavy ion reference run at 2.76\,\text{TeV} in 2013. Detectors at the LHC that could measure $\pi^{0}$ mesons during such runs are listed in the following.

The calorimeter system of the ATLAS detector \cite{ATLASJINST} reaches pseudorapidites of up to 4.9. As pointed out earlier in Sec.~\ref{intro}, the LHCf detector is located in the far forward region of the ATLAS detector, at a distance of $140\,\text{m}$ from the interaction point. At this location it is able to detect neutral particles with pseudorapidities of $|\eta|>8.4$ . The LHCb detector \cite{LHCbRef} has a single-arm forward design and covers a \prap\, range of $2<\eta<5$. While the CMS \cite{CMSRef} central detector has a similar $\eta$-coverage to that of the ATLAS detector ($|\eta|<5$), it is also equipped with a one-sided Cherenkov calorimeter called CASTOR \cite{CASTORRef} covering $-6.6<\eta<-5.2$. 

At the Tevatron proton-antiproton collider the CDF collaboration \cite{CDF:1996yrf} installed a dedicated forward upgrade for Run 2, which included forward calorimeter systems: the CDF MiniPlug detector \cite{CDFminiplug}.
It covered a \prap\, region of $3.6<|\eta|<5.1$ in proton-antiproton collisions at a \com\ energy of $1.96\,\text{TeV}$. 

During the $pp$ collision run at the Relativistic Heavy Ion Collider RHIC at a \com\ energy of $510\,\text{GeV}$ in 2017, the RHICf detector \cite{RHICf} was inserted in the forward region at one side of the STAR detector \cite{STAR} at a distance of $18\,\text{m}$ from the interaction point. RHICf is the same detector as the LHCf arm 1 detector and covers pseudorapdities of $\eta>6$. 

\begin{figure}
    \centering
    \includegraphics[scale=0.8]{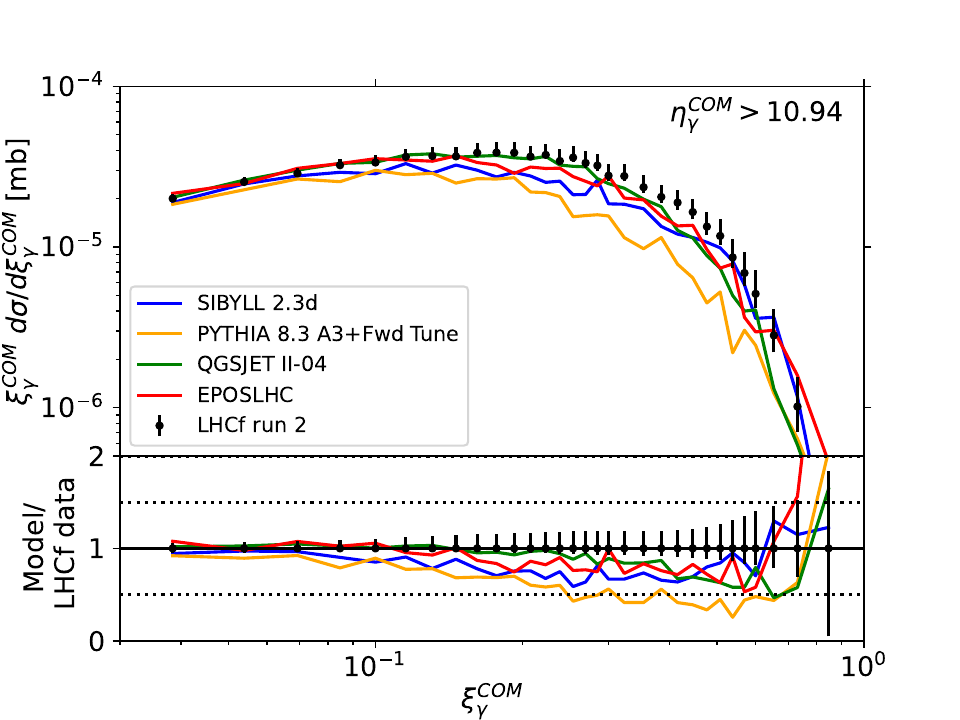}
    \includegraphics[scale=0.8]{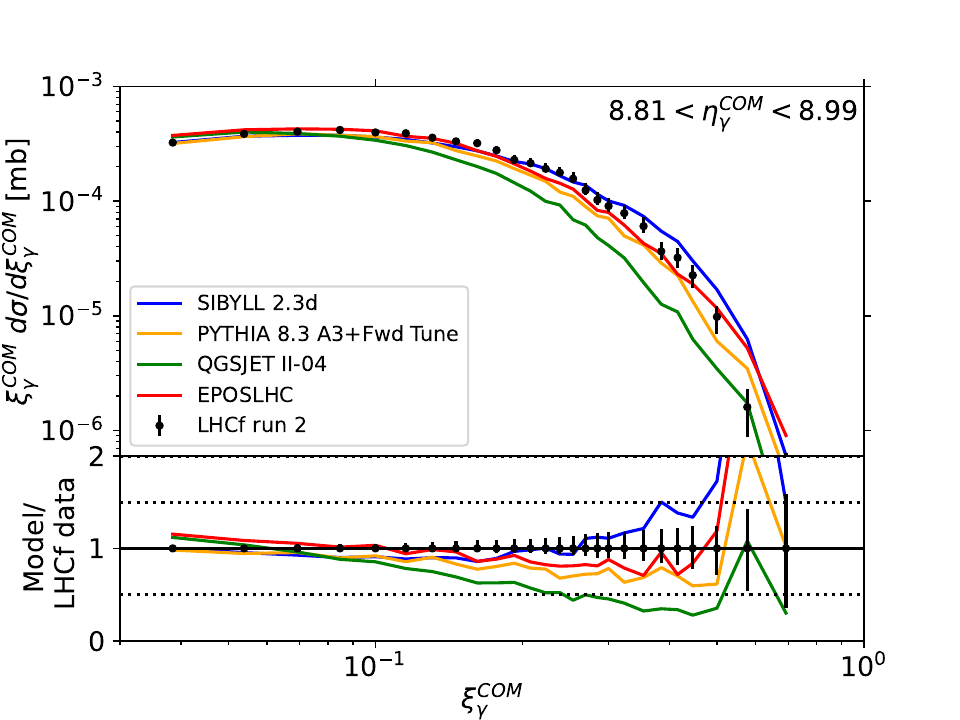}
    \caption{Comparison of the LHCf measured photon energy spectrum \cite{LHCfphotonsRun2} in $pp$ collisions at a \com\, of $13\,\text{TeV}$ in two bins of \prap\, to the corresponding interaction model predictions.}
    \label{fig:LHCfphotonComp}
\end{figure}

Fig.\,\ref{fig:EtaRegions} shows the \prap\, range of interest, and how it evolves with the \com\ energy. This dependence is in accordance with Eqn.\,\ref{eqn:etamax} and Eqn.\,\ref{eqn:etamin}, with the region between the light blue lines in Fig.~\,\ref{fig:EtaRegions} denoting this \prap\, range. This range is compared to the available datasets at the respective \com\ energy and the \prap\, range covered by each detector. It can be seen that the only detectors sensitive to $\pi^{0}$ mesons in the relevant energy range for CR backgrounds in gamma-ray analyses at IACTs are the LHCf and RHICf. 
The label ``potential'' for LHCf and RHICf data in Fig.~\ref{fig:EtaRegions} refers to data that has already been recorded but not published.

The LHCf experiment can only operate under special run conditions, i.e. a small number of simultaneous interactions per proton bunch crossing (also referred to as pileup) to reduce the received radiation dose at the detectors. It has taken data during such special $pp$ collision runs in 2010 at $7\,\text{TeV}$, in a heavy ion reference run in 2013 at $2.76\,\text{TeV}$, in 2015 at $13\,\text{TeV}$ and in 2022 at $13.6\,\text{TeV}$. The latest published results on $\pi^{0}$ measurements by LHCf \cite{LHCfPi0Run1} are based on $7\,\text{TeV}$ data and include transverse and longitudinal momentum distributions. 
The longitudinal momentum analysis was performed for one side only (only positive $\eta$) and in five bins of $p_{T}$ up to $1~\text{GeV}$, but does not include any binning in $\eta$. Similarly, LHCf measured longitudinal $\pi^{0}$ momenta at $2.76\,\text{TeV}$ in 2013 data in two bins of $p_{T}$ (also without $\eta$-binning) \cite{LHCfPi0Run1}. 
However, as was highlighted by Fig.~\ref{fig:breakdown}, the most interesting result for an application in the modeling of EM-dominated hadronic air showers is a measurement of the energy spectrum of neutral pions at high \prap, ideally as a function of \prap.

For the LHCf Run 2 data, no results on reconstructed $\pi^{0}$ mesons have been published yet but single photon energy spectra have been measured in two bins of \prap\, \cite{LHCfphotonsRun2}, $8.81<\eta(\gamma)<8.99$ and $\eta(\gamma)>10.94$ (one-sided). Although single photons do not fully represent results for neutral pions, the lack of observational data of $\pi^{0}$ leaves us no option but to compare the models to the measured single photon spectra. In Fig.~\ref{fig:LHCfphotonComp} these single photon spectra are compared to the predictions of the current interaction model versions described in Sec. \ref{Simulation}: With increasing energy the interaction model predictions deviate more strongly from the data and also from each other. This deviation is even more pronounced in the lower \prap\, bin, where the difference from the measured spectrum becomes larger than 50\% for QGSJET II-04 and Sibyll 2.3d at high photon energies. For the modeling of hadronic air showers with a large electromagnetic component it would be interesting to have a similar kind of measurement as in \cite{LHCfphotonsRun2} but for reconstructed neutral pions, and ideally with larger statistics. Analyzing the special $pp$ collision dataset that LHCf has taken in 2022 at $13.6\,\text{TeV}$ could provide such a measurement. 

The RHICf data of 2017 also falls within the defined relevant \prap\, region. However, neither a photon nor a neutral pion energy spectrum has been published yet using data taken with this instrument.

\begin{figure}[h!]
    \centering
    \includegraphics[width=0.8\linewidth]{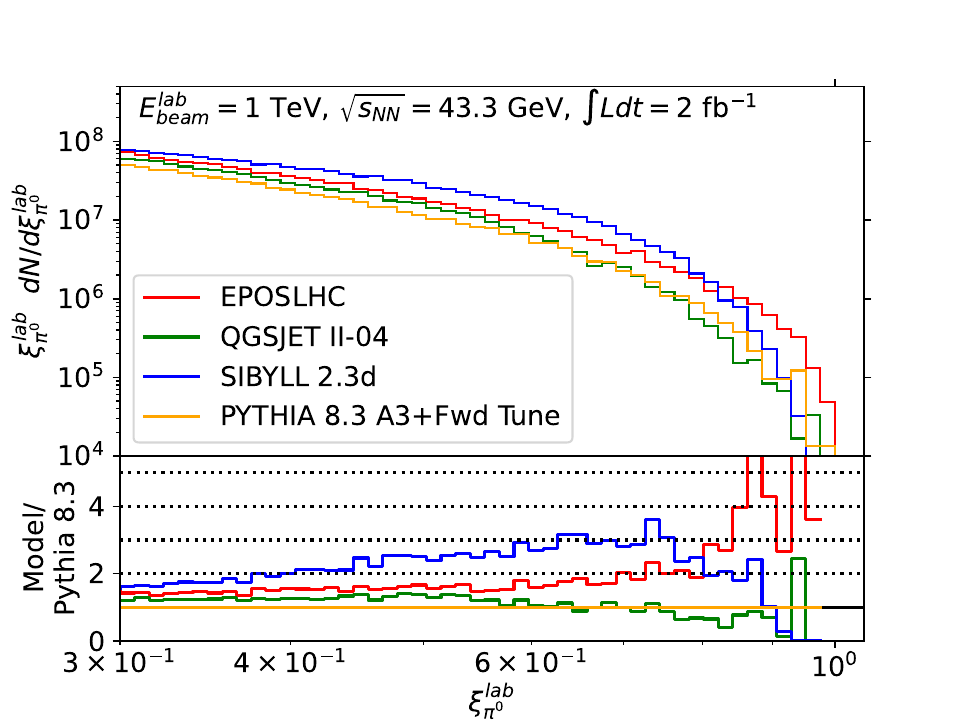}
    \caption{The predicted $\pi^{0}$ energy spectrum ($\xi_{\pi^{0}}^{\text{lab}}$) produced by a proton beam energy of $1\,\text{TeV}$ through collisions with target oxygen nuclei at rest, for four different interaction models, zoomed into the final energy decade. The differences between the models in the predicted shape of the high-energy end of the spectrum is even more pronounced than in the $pp$ collision case, which is displayed in Fig.~\ref{fig:pi0spec} (note the different scales in the $y$-axis of the ratio in the lower part of the plot).}
    \label{fig:pOGenComp}
\end{figure}


\section{Impact of Future Measurements}
\label{Discussion}

Beyond the results presented here, utilizing LHCf results from $pp$ collisions in LHC Runs 2 and 3, future results from $pp$ collisions are expected from both LHCf and RHICf. These results will offer the advantage of analyzing the $\pi^{0}$ production spectra directly. Knowledge of the specific $\eta$ bin range obtained for this data is still to be ascertained. However, the analysis of this data, decomposed into the $\eta$ range of interest here, will provide even stronger tests of the hadronic interaction model predictions for the background level of gamma-ray measurements with IACTs. Furthermore, the large (more than an order of magnitude) difference in \com\ energy for the collisions measured by RHICf and LHCf offers the unique opportunity to cross validate the tuning of a model at one \com\ energy to results at a very different \com\ energy.
It is important to note that {\sl detailed} background rate predictions for gamma-ray observations, using hadronic model simulations, also rely on accurately simulating the instrument detector efficiencies of both LHCf and IACT instruments. We have not done that in this paper. 

Although the focus in this paper has been on probing differences in the $\pi^{0}$ spectra between the hadronic interaction models in regions with large $\eta$ for $pp$ interactions, as discussed in Sec.~\ref{Air_Showers}, it is the differences of these spectra for $pO$ interactions which are of particular relevance for IACTs. Going from a description of $pp$ interactions to $p$-ion interactions further broadens the set of parameters to be considered. As an example of the corresponding broadened set of predictions for these interactions, we show in Fig.~\ref{fig:pOGenComp} 
the corresponding $\pi^0$ spectra for the case of $pO$ interactions for the four different interaction models considered. Given the scheduled $pO$ run expected in 2025, probing large $\eta$ for such collisions will become possible soon with LHCf. A separate dedicated study of the pion spectra produced in $pO$ interactions, which we leave to future work, is warranted. 

\section{Conclusions}
\label{Conclusions}

We have focused here on the dominant hadronic cosmic-ray-initiated background events which contaminate observations made by ground-based gamma-ray telescopes. These telescopes use the atmosphere as the target material for the arriving gamma rays, which subsequently initiate gamma-ray air showers. The parameters for the hadronic cosmic-ray interactions with the atmosphere, which give rise to air showers which masquerade as gamma-ray-initiated air showers, have been isolated. Specifically, the phase-space region of these hadronic interactions, whose subsequent air showers look gamma-like, has been found to have both a large energy transfer from the beam particles to the $\pi^{0}$ produced, and large pseudorapidies of these $\pi^{0}$ particles.

The current hadronic interaction models are demonstrated to vary considerably in their prediction of the $\pi^{0}$ production rates in this region of the interaction phase-space. This uncertainty on the prediction of the background rate by interaction models corresponds to the large underlying uncertainty in the interaction physics of such hadronic interactions. 

Through consideration of the phase-space region probed by recent and current collider experiments, it is demonstrated that only the LHCf and RHICf experiments are able to measure high energy forward $\pi^{0}$ production events of relevance to IACTs. A comparison of the results from LHCf 
to the current interaction models shows that no single model describes these events well.

The use of past and future LHCf and RHICf measurements of pion energy spectra can lead to improvement of event generators, which in turn will reduce the systematic uncertainty on the irreducible CR background for IACTs. Because the discrepancies between the event generator predictions are more pronounced for $pO$ collisions than for $pp$ collisions, the planned $pO$ collision run at the LHC will be of particular importance.

\begin{acknowledgements}
We would like to thank Jan Alcover-Loetsch, Marcel Konka, and Julian Rypalla for useful discussions. The authors acknowledge support from DESY (Zeuthen, Germany), a member of the Helmholtz Association HGF. 
\end{acknowledgements}

\bibliographystyle{aasjournal}
\bibliography{references.bib}

\appendix

\section{IACT Simplified Model}
\label{Appendix1}

To investigate the gamma-ray likeness of proton induced air showers (simulated at vertical incidence) simulations were created using the atmospheric density model number 22, applicable for the Malargue site in Argentina, and default extinction models, although due to the relative nature of background rejection (in comparison to gamma-ray events), the atmospheric model should not affect performance. The Cherenkov radiation produced in the showers was subsequently used in a simplified description of an array of atmospheric Cherenkov telescopes, allowing for a quick and general first order assessment of the main shower features that can be probed by such instruments. The array simulated here was comprised of nine telescopes, on a 3$\times$3 square grid, with 120\,m spacing between telescopes.

The simulation assumed that all Cherenkov photons reaching ground level within a 6\,m radius from any telescope center fall within the ``telescope" and are detected. 
The directions of the photons arriving at the telescopes were then binned into an angular array of bins centered on the pointing direction of the telescope. Photons are binned in a 40$\times$40 grid, with each bin having an angular size of 0.2$^\circ$ and scaling factors are applied to account for the telescope mirror reflectivity and camera detector efficiency
(80\% and 20\% respectively). This approach approximates the camera at the telescope focus. Gaussian smoothing with $\sigma= 0.04^\circ$ was then applied to this shower image to approximate the optical point spread function of a single Cherenkov telescope. Finally, random noise sampled from a Gaussian, with a mean of 0 (assuming that the mean noise level is subtracted in the data) and a 1~sigma width of 1 photoelectron, was added in order to simulate night sky background noise. These shower images are cleaned using a two level tailcuts cleaning method (described in \cite{HESScrab}, using the \emph{ctapipe} software \cite{linhoff_ctapipe_2023}) and characterised by the mean-scaled width and mean-scaled length parameters (MSCW, MSCL), as described in \cite{2015arXiv151005675H}.
Gamma-ray like event selection was made by requiring a minimum of 60 photoelectrons in each telescope image and a minimum of 5 images per event. Width and length cuts were then defined based on those typically used by the H.E.S.S. telescope ($-1<$MSCW$<0.7$ and $-1<$MSCL$<1$ \cite{HESScrab}). Generally these chosen cut parameters are not atypical for IACTs, and the results are insensitive to the specific values chosen. The simulation code used can be found in a GitHub repository \cite{CORSIKA_Toy_IACT}.

\section{Pseudorapidity Range at LHC energies}
\label{appendix2}

\begin{figure}
    \centering
\vspace{-2.0cm}\includegraphics[width=0.6\linewidth]{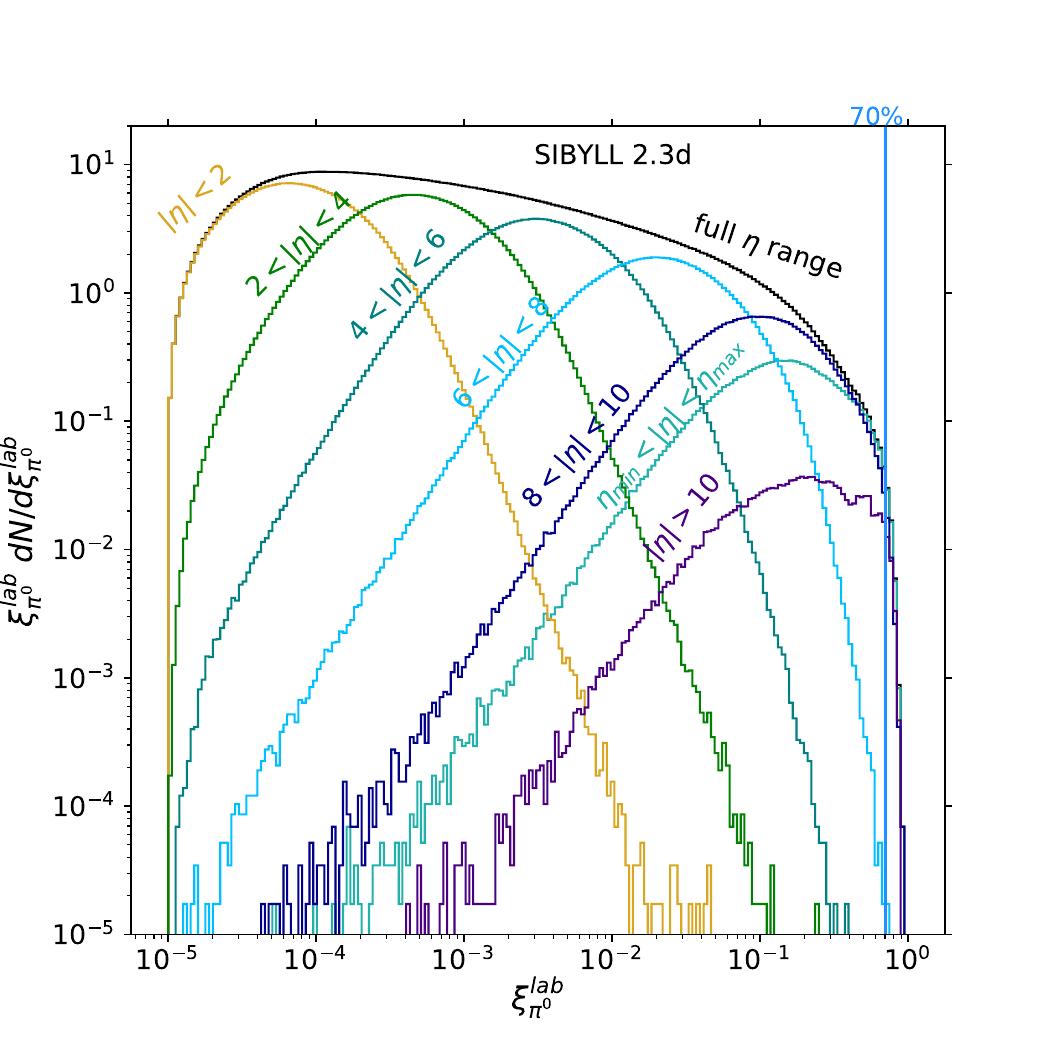}\\
\vspace{0cm}\includegraphics[width=0.6\linewidth]{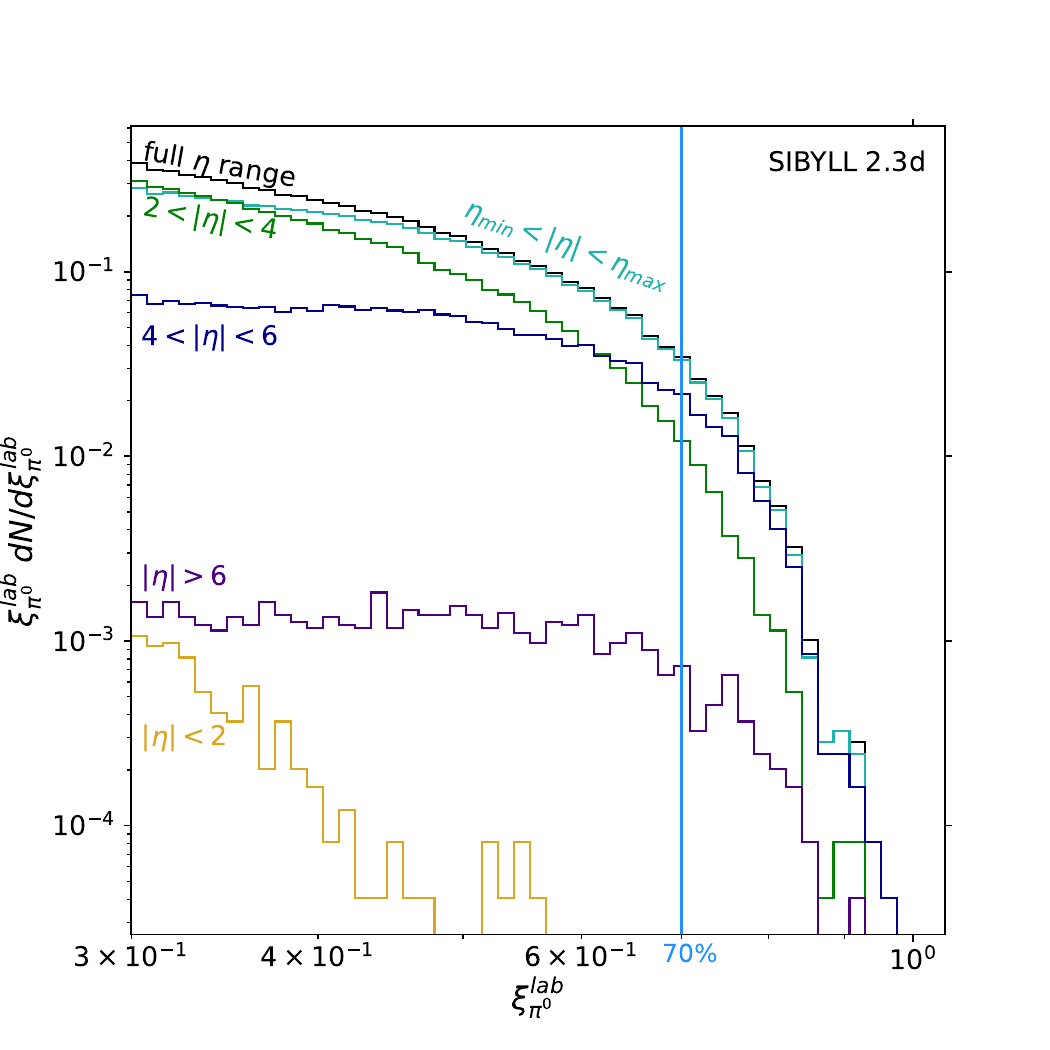}
    \caption{The $\pi^{0}$ energy spectrum produced through the collision of a $10^{17}\,\text{eV}$ proton beam with a fixed proton target (corresponding to a \com\, of $13.6\,\text{TeV}$). The full spectrum is decomposed into different \prap\, ranges (evaluated in the \com\ frame). The upper plot shows the full spectrum and the lower plot shows the zoomed in result for only the high energy fraction region. Depicted here are the predictions according to the Sibyll 2.3d event generator.
    The \prap\, range defined by equations~\ref{eqn:etamin} and ~\ref{eqn:etamax} is shown in turquoise and corresponds to $8.75<|\eta_{\pi^{0}}^{\text{COM}}|<11.5$. The 70\% percent energy fraction benchmark point is denoted by the blue vertical line.}
\label{fig:etabinsLHC}
\end{figure}

To verify that the definitions of $\eta_{\text{min}}$ and $\eta_{\text{max}}$ also show the expected behavior at higher energies, the same plots as depicted in Fig.~\ref{fig:breakdown} are produced for $pp$ collisions at a \com\, energy of $13.6\,\text{TeV}$, see Fig.~\ref{fig:etabinsLHC}. This energy corresponds to the collision energy of the LHC in Run 3 and to a cosmic-ray energy on the order of $10^{17}\,\text{eV}$. For this energy the \prap\, range as defined in Eqn.~\ref{eqn:etamax} and Eqn.~\ref{eqn:etamin} is given by $8.75 < |\eta_{\pi^{0}}^{\text{COM}}| < 11.5$. It can be observed in Fig.~\ref{fig:etabinsLHC} that this region covers the upper range of the energy spectrum ($\xi_{\pi^{0}}^{\text{lab}}>70\%$) fully. We conclude that these definitions for $\eta_{\text{min}}$ and $\eta_{\text{max}}$ serve as a good approximation for the region of interest for IACT background modeling for different orders of magnitude in energy. 

The low energy cutoff in the spectrum observed in Fig.~\ref{fig:etabinsLHC} originates from neutral pions produced at rest in the \com\ frame. The difference in the position of the cutoff in this figure compared to that in Fig.~\ref{fig:breakdown} comes from the difference in \com\ energy for the collision.

\end{document}